\documentstyle[12pt, aaspp4]{article}

\newcommand{\lsun}{$\log L/L_{\odot}\,$}
\newcommand{\msun}{$M/M_{\odot}\,$}

\begin{document}

\title{Linear nonadiabatic properties of SX Phoenicis variables}

\author{P. Santolamazza\altaffilmark{1}, M. Marconi\altaffilmark{1}, 
G. Bono\altaffilmark{2}, F. Caputo\altaffilmark{2}, 
S. Cassisi\altaffilmark{3} and R. L. Gilliland\altaffilmark{4} }

\lefthead{Santolamazza et al.}
\righthead{SX Phoenicis}

\altaffiltext{1}{Osservatorio Astronomico di Capodimonte, Via Moiariello 16,
80131 Napoli, Italy; patrizia@na.astro.it; marcella@na.astro.it}
\altaffiltext{2}{Osservatorio Astronomico di Roma, Via Frascati 33,
00040 Monte Porzio Catone, Italy; bono@coma.mporzio.astro.it,
caputo@coma.mporzio.astro.it}
\altaffiltext{3}{Osservatorio Astronomico di Collurania, Via M. Maggini,
64100 Teramo, Italy; cassisi@astrte.te.astro.it}
\altaffiltext{4}{Space Telescope Science Institute, 3700 San Martin Drive, 
Baltimore, MD 21218; gillil@stsci.edu}

\begin{abstract}

We present a detailed linear, nonadiabatic pulsational scenario for 
oscillating Blue Stragglers (BSs)/SX Phoenicis (SX Phe) in Galactic 
Globular Clusters (GGCs) 
and in Local Group (LG) dwarf galaxies. The sequences of models 
were constructed by adopting a wide range of input parameters 
and properly cover the region of the HR diagram in which these 
objects are expected to be pulsationally unstable. Current calculations
together with more metal-rich models already presented by Gilliland et al. 
suggest that the pulsation properties of SX Phe are partially affected 
by metal content. In fact, the pulsation periods for the first three 
modes are marginally affected when moving from Z=0.0001 to 0.006, 
whereas the hot edges of the instability region move toward cooler 
effective temperatures by approximately $300\div500$ K.   

The inclusion of a metallicity term in the Period-Luminosity-Color 
(PLC) relations causes a substantial decrease in the intrinsic 
scatter and in the individual error of the coefficients. This supports 
the result recently brought out by Petersen \& Christensen-Dalsgaard  
for $\delta$ Scuti stars. Moreover, we find that the discrepancy  
between our relation and similar theoretical and empirical relations  
available in the literature is typically smaller than 5\%. 
The comparison between theory and observations in the $M_V - \log P$ 
plane as well as in the luminosity amplitude $- \log P$ plane does not help 
to disentangle the long-standing problem of mode identification among 
SX Phe stars. 
However, our calculations suggest that the secular period change seems 
to be a good observable to identify the pulsation mode of cooler SX Phe 
variables. 

Together with the previous models we also constructed new sequences of 
models by adopting selected effective temperatures and luminosities 
along two evolutionary tracks characterized by the same mass value and metal 
content (\msun=1.2, Z=0.001) but different He contents in the envelope, 
namely Y=0.23 and Y=0.30. The He content in the latter track was 
artificially enhanced soon after the central H exhaustion to mimic,  
with a crude approximation, the collisional merging between two stars. 
Interestingly enough, we find that the He-enhanced models present 
an increase in the pulsation period and a decrease in the total kinetic 
energy of the order of 20\% when compared with the canonical ones. 
At the same time, the blue edge of the fundamental mode for the 
He-enhanced models is approximately 1,000 K cooler than for canonical 
ones. Moreover, we find that the secular period change for He-enhanced 
models is approximately a factor of two larger than for canonical ones. 
According to this evidence we suggest that the pulsation properties 
of SX Phe can be soundly adopted to constrain the evolutionary history 
of BSs, and in turn to single out the physical mechanisms which trigger 
their formation.  
\end{abstract}

{\em Subject headings:} $\delta$ Scuti -- stars: blue stragglers -- 
stars: evolution -- stars: oscillations -- stars: variables: other

\pagebreak
\section{Introduction}

During the last few years both ground based and HST photometric 
observations of Open Clusters (Ahumada, \& Lapasset 1995; 
Dinescu et al. 1996; Landsman et al. 1998; Mazur et al. 1999b), 
GGCs (Auriere, Lauzeral, \& Ortolani 1990; 
Buonanno et al. 1994; Ferraro, Fusi Pecci, \& Bellazzini 1995; 
Walker \& Nemec 1996; Kaluzny et al. 1997; Guarnieri et al. 1998; 
Borissova et al. 1999; Ferraro et al. 1999; Piotto et al. 1999), 
and dwarf Spheroidal galaxies (dSphs, Mateo et al. 1995; 
Grillmair et al. 1998; Monkiewicz et al. 1999) brought out the 
evidence that BSs are more a widespread feature than an anomaly 
in these stellar systems. 
The BSs are systematically bluer than cluster main-sequence (MS) 
stars and typically cover a range of two magnitudes above the 
Turn-Off (TO). Therefore BSs in these stellar systems mimic a 
population of MS and post-MS stars more massive than the TO stars. 
The observational scenario of these intriguing objects is further 
enriched by the fact that BSs cross the Cepheid instability strip, 
and therefore a fraction of them, roughly the 30\%, oscillate both 
in radial and/or nonradial modes (Gilliland et al. 1998, hereinafter G98). 
The oscillating BSs after the discovery in $\omega$ Cen (Niss 1981) 
were christened SX Phoenicis variables (Nemec \& Mateo 1990).  
 
The BSs are the crossroad of paramount theoretical and observational 
efforts, since we still lack a firm understanding of the physical processes 
that govern their origin and their subsequent evolution. Even though 
several mechanisms have been proposed to explain the occurrence of BSs 
in GCs, we have not yet understood how the dynamical history of the 
cluster cores affects the formation of these objects. 
The most popular mechanism suggested for the formation of BSs is  
the mass accretion of a MS star caused either by the mass 
transfer/coalescence between different components in primordial 
binary systems (binary merging) or the direct collision of two or 
more stars (collisional merging) in high density regions (Bailyn 1995; 
Sandquist, Bolte, \& Hernquist 1997, hereinafter SBH97; 
Sills \& Bailyn 1999). These two 
different hypotheses have been further strengthened by the discovery 
of contact, semidetached, and detached eclipsing binaries in several 
GCs (Mateo et al. 1990; Kaluzny, \& Krzeminski 1993; Edmonds et al. 1996; 
Rubenstein, \& Bailyn 1996; Yan \& Reid 1996; Mazur, Kaluzny, \& 
Krzeminski 1999a, and references therein). 

Moreover, dynamical calculations suggest that the binding energy in 
primordial binaries is a key source to prevent, if not halt, the core 
collapse and/or to trigger the expansion during the post-core collapse
phases. As a consequence, if BSs are formed via a binary merging, the 
luminosity function of BSs can constrain the energy supplied by these 
objects (SBH97). On the other hand, if BSs are the progeny of star-star, 
binary-star, or binary-binary interactions, then the luminosity functions 
of clusters characterized by different central densities can supply tight 
constraints on the cross section of stellar interactions (SBH97).  

In order to properly identify binary and collisional BSs, Bailyn (1992) 
suggested a photometric criterion. In fact, hydrodynamical calculations 
by Benz \& Hills (1992) show that the remnants of two colliding stars   
are characterized by a higher He content in the envelope than the 
remnants of binary merging. As a consequence, the BSs formed via 
collisional merging should be simultaneously brighter and bluer 
than the ones formed via binary merging. The difference in the 
chemical profile would also imply a substantial change in the 
evolutionary tracks and in the luminosity functions of the two 
BSs groups (Bailyn \& Pinsonneault 1995; SBH97). 
A further interesting observable to split binary and collisional 
BSs has been suggested by Shara, Saffer, \& Livio (1997) and by 
Ouellette, \& Pritchet (1998) who noted that the remnants of 
collisional mergings should be, due to the conservation of 
angular momentum, rapidly rotating stars.   

Although these selection criteria appear to be based on straightforward 
empirical stellar properties and an unprecedented number of BSs stars have 
been detected by HST observations in the dense cores of GGCs we still 
lack a clear understanding on the intimate nature of these objects.  
In fact, recent deep photometric data support the evidence that BSs 
present a bimodal radial distribution (Bailyn, \& Pinsonneault 1995; 
Ferraro et al. 1997; Saviane et al. 1998). 
In this scenario the BSs located close to cluster center were formed 
via stellar collisions, whereas the BSs located in the outskirt of 
the cluster formed via binary merging. However, no general consensus 
has been reached yet, and indeed Sigurdsson et al. (1994) suggested 
that the entire sample of BSs in M3 formed via stellar collisions, 
while Sills, \& Bailyn (1999) suggested that in this cluster the BSs 
centrally located are the progeny of two distinct populations.  

In view of these unquestionable facts, we suggest to investigate the 
pulsation behavior of SX Phe variables to constrain the physical mechanisms 
that trigger the formation  and evolution of BSs. In fact, it has been 
recently shown (Petersen \& Christensen-Dalsgaard 1996; G98; Petersen
\& H\"og 1998a) that the comparison between predicted and observed period 
ratios can supply useful constraints on the pulsation masses of mixed-mode 
variables. 
Interestingly enough, G98 found a good agreement between pulsational and 
evolutionary masses for 4 mixed-mode variables and by combining the two 
estimates it turned out that the stellar masses of these objects range 
from $1.3\pm0.1$ to $1.6\pm0.2$ $M_\odot$.  
Although this approach presents several advantages, the theoretical 
predictions currently adopted in the literature to constrain the 
pulsational behavior of SX Phe are only based on linear adiabatic 
and nonadiabatic models. Therefore the comparison between theory and 
observations was limited to period and period ratios. At the same 
time, the mode identification was based on linear growth rates 
instead of on nonlinear limit cycles. This is the second paper of 
a series devoted to a detailed theoretical and observational analysis 
of SX Phe variables in stellar systems and in the Galactic field. 

The main aim of this investigation is to present new linear, 
nonadiabatic pulsation models for metal-poor (Z=0.0001 and Z=0.001) 
SX Phe and to compare theoretical predictions with observational 
data available in the literature. 
In \S 2 we discuss current linear, nonadiabatic models and present 
a detailed analysis of the dependence of pulsation properties on 
metallicity. 
In \S 3.1 we supply the analytical relations for the first 
three radial modes connecting the pulsation period to stellar mass, 
luminosity, effective temperature, and metallicity. 
Both the PLC and the PLCZ relations and their intrinsic accuracies 
are detailed in \S 3.2, together with the comparison with similar
theoretical and empirical relations available in the literature. 
The comparison between the predicted blue edges for the first three 
radial modes with current empirical data on SX Phe in GGCs, in dwarf 
galaxies, and in the Galactic field is presented in \S 4. In this 
section the problems that are still hampering the mode identification 
among SX variables are also briefly discussed. 
The dependence of modal stability, and pulsation periods on the He 
content is presented in \S 5. The impact that the comparison between 
empirical and theoretical observables could provide to disentangle 
the evolutionary history of BSs are also outlined in this section. 
The main results of this investigation are summarized in \S 6.

\section{Linear Nonadiabatic Models}

The input physics, physical and numerical assumptions adopted to construct
linear, nonadiabatic, radiative models for SX Phe variables were already
addressed in Bono et al. (1997, hereinafter B97) and in G98. Therefore they
are not discussed in detail here.
In order to supply a comprehensive theoretical framework for SX Phe in 
GGCs we constructed several sequences of models by adopting a fixed He 
content (Y=0.24) and two different metal abundances, namely Z=0.0001 
and Z=0.001.  
For each given chemical composition we adopted three different mass 
values (\msun=1.0, 1.2, 1.4) and luminosity levels ranging from 
\lsun = 0.60 to \lsun =1.4. The models cover a wide temperature range 
($6,000 \le T_e \le 8,800$ K) and for each individual model we investigated 
the modal stability of the first three radial modes. 

Tables 1-3\footnote{Table 1 is published in the printed edition, while 
Tables 2 and 3 are only available in the on-line edition.} summarize the 
input parameters and selected linear observables
for the entire set of models we constructed. From left to right the 
columns give: luminosity, effective temperature (K), radius (cm), 
period (day), pulsation constant (d), kinetic energy (ergs), and 
growth rate. The third column of Tables 2 and 3 lists the first overtone 
(FO) and the second overtone (SO) to fundamental (F) period ratios 
respectively.  
In agreement with the assumption adopted by G98 the inner boundary was 
fixed at a relative distance from the stellar center roughly equal to 
10 \% of the photospheric radius. This choice ensures that the 
envelope mass throughout the instability strip ranges from 40 to 80\% 
of the total mass. 
The mass ratios between consecutive zones and the spatial resolution
across the envelope were also fixed according to G98. For temperatures
higher than 10,000 K we adopted OPAL radiative opacities (Iglesias \&
Rogers 1996), while for lower temperatures we adopted molecular opacities
by Alexander \& Ferguson (1994). The procedure adopted to interpolate the
opacity tables is described in Bono, Incerpi, \& Marconi (1996). 
The small values attained by the growth rate explains why up to now 
the lengthy calculations necessary to investigate the nonlinear 
behavior of high gravity variables have been only partially tackled 
(Bono et al. 1997).  

Figure 1 shows the opacity and the adiabatic exponent $\Gamma_3$ 
(left panels) as a function of logarithmic temperature for three 
different models centrally located in the instability strip. 
These models were constructed by adopting the same stellar mass 
(\msun=1.2), luminosity (\lsun=1.05), effective temperature 
($T_e=7250$ K), and He content (Y=0.24) but three different 
metal abundances, namely 0.0001, 0.001, and 0.006. Data plotted 
in this figure show quite clearly that these structures present 
a mild dependence on metal content, and indeed a decrease of 
almost 1.5 dex in [Fe/H] causes a decrease of the order of 
20\% in the Z-bump opacity ($\log T \approx5.15$)
and a negligible change in both HeII ($\log T \approx4.65$) 
and H plus HeI ($\log T\approx4.18$) opacity peaks. The adiabatic 
exponent also shows a negligible dependence on Z, and indeed the 
three models attain, in this temperature interval, almost identical 
values. The mild dependence on Z is also supported by F periods 
which increase by approximately 3\% when moving from Z=0.0001 
to 0.006. 

The total works plotted in the right panels show, as expected (Bono,
Marconi \& Stellingwerf 1999a), that an increase in the metal content 
from 0.0001 to 0.006 causes a decrease in the pulsation 
destabilization. In fact, the total work of the F mode decreases from 
$7.15\times10^{-6}$ (Z=0.0001) to $3.80\times10^{-6}$ (Z=0.006). 
Even though the Z-bump in the more metal-rich model (dashed line) is
larger than in the metal-poor ones (solid line) the driving across
this region is smaller in the former model that in the latter ones.
This effect might be due to the fact that the opacity derivatives in
the more metal-rich model are smoother than in the metal-poor ones.
The FO mode shows a similar behavior, and indeed the pulsation periods
increase by less than 1\%, while the total work decreases from
$7.97\times10^{-5}$ (Z=0.0001) to $5.02\times10^{-5}$ (Z=0.006).
We refer the reader to G98 for a detailed discussion on the role
played by the Z-bump and by the location of radius, luminosity and
temperature nodes in the destabilization of overtone pulsators. 
Note that even though the dependence of the SX Phe pulsation 
properties on metallicity is marginal, an increase in the metal 
content causes a shift of the fundamental blue edge (FBE) toward 
cooler effective temperatures by approximately 200 K at \lsun=1.0 
and 500 K at \lsun=0.7. In this luminosity range first and second 
overtone blue edges move toward lower effective temperatures by 
roughly 400 K. This suggests that metal-poor BSs are, at fixed 
input parameters, more enthusiastic pulsators than metal-rich 
ones.  

The small values attained by the quoted total works together with 
the values listed in column (7) of Tables 1-3 confirm the strong 
adiabatic nature of SX Phe brought out by G98. Current results 
also suggest that at fixed input parameters a decrease in metallicity  
causes a further decrease in the total work, and in turn in the 
adiabatic nature of these objects.  
The normalized total work is the analog of the growth rate, i.e. 
the stability parameter which supplies a necessary condition 
for the modal stability. A glance at the models which are 
unstable in the first three modes suggests that the SO total 
works are higher than for FO and the latter ones higher than 
for F pulsators. This finding confirms that even in the metal-poor 
regime higher overtones are more destabilized than the F mode (G98). 
Moreover, the evidence that at fixed input parameters 
the total kinetic energy decreases by approximately a factor of 5 
between F and FO and by a factor of 2.5 between first and second 
overtone makes the probability to detect overtone pulsators among 
SX Phe higher than for F ones. 
Finally, we mention that a decrease in the metal content causes on 
average an increase in the period ratio. In fact, the $P_1/P_0$ 
increases from 0.77 to 0.79 when moving from Z=0.006 to Z=0.0001. 
The difference among higher overtones is smaller, and indeed 
the same change in metallicity causes an increase in $P_2/P_1$ 
from 0.81 to 0.82. The small changes in the period ratios of 
consecutive models listed in Table 2-3 are caused by marginal 
variations in the total mass of the envelope.

\section{Systematic pulsation properties}

\subsection{Pulsation periods}

Recent optical and UV data collected with HST allowed the identification 
and the follow-up of exotic X-ray sources in the core of GCs (Grindlay 1999). 
Quite often these objects present luminosity variations on time scales which 
range from few minutes to 17 hours. As a consequence, they partially cover 
the period range of SX Phe (Deutsch, Margon, \& Anderson 2000).  
In order to supply the pulsation relations -i.e. the analytical relations 
which connect the pulsation period to stellar mass, luminosity, and 
effective temperature- for the first three radial modes, we estimated 
the least square fit to the large grid of models listed in Tables 1-3. 
However, the comparison between the periods based on the new relations 
and the periods based on the analytical relations derived by G98 for 
Z=0.006 presents a period difference that for the second overtone 
is the of order of 0.015 in $\log P$. Therefore, to derive pulsation 
relations which cover a wide metallicity range ($0.0001 \le Z \le 0.006$) 
we performed once again  the fit by including the G98 models. 
The analytical relations we find are the following: 

$\log P (F)= 9.854(\pm0.003) -0.434(\pm0.005)\cdot \log M +0.743(\pm0.001)\cdot \log L$ 

\hspace*{2cm} $-3.016(\pm0.005)\cdot \log T_e + 0.0023(\pm0.0004)\cdot \log Z$ ~~~~~~~~~~~~ $\sigma=0.003$

$\log P (FO)= 9.800(\pm0.004) -0.401(\pm0.006)\cdot \log M +0.736(\pm0.001)\cdot \log L$

\hspace*{2cm} $-3.034(\pm0.006)\cdot \log T_e + 0.0043(\pm0.0005)\cdot \log Z$ ~~~~~~~~~~~~ $\sigma=0.004$

$\log P (SO)= 9.620(\pm0.004) -0.382(\pm0.006)\cdot \log M +0.725(\pm0.001)\cdot \log L$

\hspace*{2cm} $-3.011(\pm0.005)\cdot \log T_e + 0.0081(\pm0.0004)\cdot \log Z$ ~~~~~~~~~~~~ $\sigma=0.004$

where M and L are in solar units, $\sigma$ is the standard deviation and 
the other symbols have their usual meaning. Note that the sign of the 
metallicity term supplies a direct explanation of the empirical evidence 
recently brought out by Rodriguez \& Lopez-Gonzalez (2000, hereinafter RLG00) 
that the mean period of SX Phe in GGCs and in LG dwarfs scales with the 
metallicity.  
The accuracy of the new analytical relations is further strengthened by 
data plotted in Figure 2 that shows the difference between computed and 
analytical periods for the three selected metallicities. 
In fact, the discrepancy for the first three modes is, over the entire 
period range, typically smaller than 0.005 dex.  

Finally, we mention that current models have been constructed by assuming
that radiation is the main flux carrier throughout the envelope. This is
a plausible assumption for hot models but could introduce systematic
uncertainties in the coolest ones. In order to supply a quantitative
estimate of this effect we evaluated the difference between the periods
for $\delta $ Scuti stars predicted by B97 on the basis of both linear,
nonadiabatic, radiative models and full amplitude, nonlinear, convective
models (see their Tables 1 and 2). We find that the difference for
the first three radial modes ranges from approximately $9\times 10^{-4}$
$2\times 10^{-3}$. As expected the nonlinear, convective periods are
systematically shorter than the linear, radiative ones (Bono \&
Stellingwerf 1994). This difference might introduce spurious effects in
the period ratios predicted by radiative models but it has negligible
effects on individual periods.
On the other hand, the assumption of radiative transport marginally
affects the linear predictions on the blue edges. In fact, they are
typically located at effective temperatures hotter than 7,500 K, and
therefore radiation is the main flux carrier throughout the envelope.
The comparison between linear and nonlinear blue edges predicted
by B97 is hampered by the occurrence of mixed-mode pulsators across
the edges.

\subsection{Theoretical Period-Luminosity-Color relations}

In two recent detailed investigations McNamara (1997,2000) 
found that High-Amplitude $\delta$ Scuti (HADS, i.e. variables 
with visual amplitudes larger than $0.15\div0.30$ mag) and SX Phe stars 
obey to a Period-Luminosity (PL) relation. This finding was 
supported by Petersen \& Christensen-Dalsgaard (1999, hereinafter PCD99) 
who also found that the intrinsic dispersion of the theoretical PL 
relation can be significantly reduced by including both a color 
and/or a metallicity term, i.e. by adopting a PLC or a PLCZ relation. 
It is worth mentioning that the previous authors, according to 
empirical evidence on the width of the instability strip, found 
that this outcome holds not only for HADS but also for the broad 
family of $\delta$ Scuti stars. 

To assess whether SX Phe pulsators follow the same trend we estimated 
the least square solutions for the sample of models listed in Table 1-3 
and for G98 models. The analytical PLC and PLCZ relations for the first 
three radial modes we find are the following:

\begin{tabular}{lllll}
$M_{bol}(F) \;\;\;=$& 41.06&$ -3.69\, \log P$&$-11.17\,\log T_e$&($\sigma$=0.08)\\
                &$\pm0.08$& $\pm0.03$       &  $\pm0.18$ &       \\
$M_{bol}(FO)\,=$& 39.79&$ -3.68\, \log P$&$ -10.93\,\log T_e$& ($\sigma$=0.09)\\
                &$\pm0.09$& $\pm0.03$       &  $\pm0.17$ &      \\
$M_{bol}(SO)\,=$& 38.20&$ -3.67\, \log P$&$ -10.61\,\log T_e$& ($\sigma$=0.09)\\
                &$\pm0.09$& $\pm0.02$       &  $\pm0.19$ &      \\
\end{tabular}

where the pulsation periods are in days, the effective temperatures
in K, and the bolometric magnitudes were derived by assuming  
$M_{bol}({\odot})=4.62$ mag. The $\sigma$ values given at the end 
of each relation is the intrinsic dispersion. The PLCZ relations 
are the following:  

\begin{tabular}{llllll}
$M_{bol}(F) \;\;\;=$& 41.60&$ -3.73\, \log P$&$ -11.37\,\log T_e$& $-0.07\,\log Z$& ($\sigma$=0.06)\\
                &$\pm0.06$& $\pm0.02$       &  $\pm0.14$ & $\pm0.01$ &        \\
$M_{bol}(FO)\,=$& 40.87&$ -3.73\, \log P$&$ -11.30\,\log T_e$& $-0.09\,\log Z$& ($\sigma$=0.06)\\
                &$\pm0.06$& $\pm0.02$       &  $\pm0.12$ & $\pm0.01$ &        \\
$M_{bol}(SO)\,=$& 40.22&$ -3.76\, \log P$&$ -11.24\,\log T_e$& $-0.10\,\log Z$& ($\sigma$=0.06)\\
                &$\pm0.06$& $\pm0.02$       &  $\pm0.12$ & $\pm0.01$ &        \\
\end{tabular}

Previous results support the finding brought out by PCD99, and 
indeed not only the errors on the coefficients but also the formal 
scatter of the PLCZ relations are systematically smaller than for 
the PLC relations. In particular, we find that the $\sigma$ 
values in the former relations decrease by a factor ranging 
from 30\% (F) to 50\% (FO, SO). 
Note that current relations were derived by including all 
unstable models i.e. no temperature selection was performed.  

Figure 3 shows the fractional difference in the bolometric magnitude 
for the fundamental mode between 
our PLCZ relation and the PLCZ relations derived by PCD99 on the basis of
theoretical models (triangles) and empirical data (circles) collected 
by McNamara (1997). The comparison was performed at fixed metallicity
Z=0.006, since the metal content of the models constructed by PCD99  
range from  Z=0.005 to Z=0.02, while the metal abundances of 
HADS collected by McNamara range from Z=0.0001 to Z=0.02.  
The agreement between the three relations is remarkable, since the 
discrepancy is typically smaller than 5\% over the entire period 
range. This finding leads further support to the evidence that both 
SX Phe and $\delta$ Scuti can be safely adopted to estimate stellar 
distances. Moreover, previous PLCZ relations suggest that SX Phe 
pulsating in the first or in the second overtone could be good 
standard candles as well. In fact, the $\sigma$ values of their 
PLCZ relations are, within the errors, identical to the fundamental 
one. 

The agreement in the sign of the metallicity term in both 
theoretical and empirical relations brings out a key result:
{\em metal-poor SX Phe and $\delta$ Scuti stars are at fixed period  
and effective temperature fainter than metal-rich ones}. 
This finding is interesting not only because observations and 
theoretical predictions span a three dex metallicity range but 
also because it supports recent empirical and theoretical 
predictions based on the PLC (Bono et al. 1999a,b; Groenewegen 2000) 
and on PLCZ (Bono \& Marconi 1999) of classical Cepheids\footnote{
Note that in the PL plane the dependence on metallicity is exactly 
the contrary i.e. at fixed period metal-poor Cepheids are brighter 
than metal-rich ones.}. In fact, as suggested by Fernie (1992), and 
more recently supported by McNamara (1997), $\delta$ Scuti and classical 
Cepheids seem to follow the same PLC relations.

\section{Comparison between theory and observations}

Up to now 122 SX Phe have been detected in 18 GGCs and 27 in two LG 
dwarf galaxies (Carina, Sagittarius). The metallicity of these 
objects ranges from [Fe/H]=-0.70 (47~Tuc) to -2.4 (NGC~5053). The reader 
interested in an overview of SX Phe observational properties is referred 
to the recent paper by RLG00. In order to derive a large data sample of both 
cluster and field SX Phe we selected among the field objects available 
in the literature the stars with [Fe/H]$\le -0.50$, since the metallicity 
of our models ranges from Z=0.0001 to Z=0.006. We end up with 12 field 
SX Phe stars. 

The pulsation period, the absolute magnitude and the visual amplitude 
for the entire sample are listed in Table 4 (available in the on line
edition). Note that our sample of SX Phe in GGCs is not identical 
to the RLG00 one. We did not include the variables G172 and H2 detected 
in M~4 and M~71, since they present quite different color than the bulk 
of SX Phe stars. We also excluded 4 SX Phe in NGC~6397, since their 
period is uncertain, but we included the SX Phe recently discovered by 
Mochejska, Kaluzny, \& Thompson (2000) in E~3. At the same time, we also 
excluded 4 SX Phe in Carina, since only B magnitudes are available for 
these objects. We end up with a sample of 117 SX Phe in GGCs and with 
a sample of 23 SX Phe in dwarf galaxies.  

The periods in the sample of SX Phe we collected range from 45 minutes 
to roughly 4 hours ($\log P=-1.54 \div -0.8$) and V amplitudes $A_V$ 
from 0.02 to 0.90 mag. In order to shed more lights on the long-standing 
problem of mode identification we plotted the entire sample of SX Phe 
in the Bailey diagram, i.e. $A_V$ vs period. We find that these 
observables are useless to disentangle this problem, since over the 
entire period range the visual amplitudes changes from few hundredths 
up to tenths of magnitudes. The same outcome applies if we split the 
sample by adopting different metallicity bins. This suggests that in 
the Bailey diagram no clear separation emerge between SX Phe pulsating 
in the fundamental or in higher overtones. 

Even though it has been generally assumed that large luminosity 
amplitudes ($A_V\ge 0.15\div 0.3)$ are a key feature of fundamental 
pulsators (McNamara 1995,1997) both empirical and theoretical evidence 
suggest that this criterion could be affected by selection biases. 
In fact, current nonlinear, convective models of $\delta$ Scuti stars
(Bono et al. 1997) support the evidence that overtone pulsators 
present large luminosity
amplitudes close to the hot edge of the instability strip, while 
fundamental pulsators present small amplitudes close to the cool 
edge. The double-mode SX Phe V2 detected in 47~Tuc (G98) presents 
a V amplitude equal to 0.051 in the fundamental mode. This means 
that a selection criterion only based on the pulsation amplitude 
could misidentify low-amplitude fundamental pulsators. 
Moreover, as already noted by G98, the detection of low-amplitude, 
long-period SX Phe could be affected by selection effects. 

We also note that SX Phe in Carina are characterized on average by 
larger luminosity amplitudes $A_V\approx0.55$ mag when compared with the 
amplitudes of SX Phe in GGCs. However, this sample could be affected 
by potential selection effects, since the photometric accuracy at the 
typical mean magnitude of SX Phe in Carina was of the order of 0.05 mag
(Mateo et al. 1998). On the other hand, the three SX Phe detected in 
Sagittarius present normal amplitudes, but unfortunately their membership 
has not been confirmed yet. As a consequence, current data do not allow 
us to assess on a firm basis what is typical from what is peculiar, 
and in particular whether SX Phe in dwarf galaxies and in GGCs share 
the same properties.

To assess whether linear, nonadiabatic models can help to disentangle 
this thorny problem the theoretical blue (hot) edges of the instability 
strip for the first three radial modes were transformed into the 
observational plane by adopting the bolometric corrections provided 
by Castelli, Gratton \& Kurucz (1997a,b). Owing to the mild dependence 
of the blue edges on the metal content,  we derived by adopting the 
entire set of models -i.e. current plus G98- the following analytical 
relations:  

\begin{tabular}{llll}
$M_V^B(F) \;\;\;=$& -1.32 &$ -3.05\, \log P$& ($\sigma$=0.07)\\
                &$\pm0.06$& $\pm0.05$ &       \\
$M_V^B(FO)\,=    $& -1.88 &$ -3.13\, \log P$& ($\sigma$=0.05)\\
                &$\pm0.05$& $\pm0.04$ &      \\
$M_V^B(SO)\,=    $& -2.43 &$ -3.26\, \log P$& ($\sigma$=0.05)\\
                &$\pm0.06$& $\pm0.04$ &      \\
\end{tabular}

where the symbols have their usual meaning. 
Figure 4 shows the comparison in the Magnitude-Period diagram between 
current empirical data in GGCs (top) and in dwarf galaxies (bottom) 
with the blue edges for the first three radial modes. Distance moduli 
and metallicities for GGCs were taken from Harris (1996) and Carretta 
\& Gratton (1997), while for 
Carina and Sagittarius we adopted the values suggested by Mateo (1998). 
A glance at the data plotted in this figure clearly shows that the bulk 
of SX Phe (roughly 70\%) are distributed in the so-called "OR-region" 
i.e. the region of the instability strip in which two or more modes 
are simultaneously excited. Therefore this theoretical scenario does 
not allow us to firmly identify the mode of these variables. 
However we note that, within current uncertainties on GGC distances, 
the variables located above the FBE are almost certainly overtone 
pulsators. The location of blue edges is only marginally affected by 
nonlinear effects and therefore the nonlinear blue edges should only 
be slightly shifted toward cooler effective temperatures.  
Moreover, the four brightest variables among the SX Phe identified 
in 47~Tuc are also located above the FBE, in particular the double-mode 
variable V15 is almost certainly pulsating in overtones higher than the 
second one. This evidence supports the mode identification suggested 
by G98 for this object. The same outcome applies for V2.  

Figure 5 shows the distribution of field SX Phe in the $M_V-\log P$ diagram. 
The comparison with the theoretical edges brings out the same behavior 
we already found for SX Phe in GGCs and in dwarfs. Three fundamental/first 
overtone double-mode variables have been identified to date among field 
SX Phe, namely BL Cam (Hintz et al. 1997a), SX Phe (Nemec, Nemec-Linnell, 
\& Lutz 1994), and AE UMa (Hintz et al. 1997b). Note that among these 
variables the main pulsation mode, i.e. the mode with the largest 
luminosity amplitude, is the fundamental one. Our results seem to 
support these identifications, since these objects are located 
within the errors across the fundamental blue edge.  
Due to the well-known correlation between the pulsation periods 
and metallicity we also tested whether the metal content plays  
any role in the mode identification. We split the SX Phe in GGCs 
in two metallicity bins i.e. $[Fe/H]\le -1.6$ and $[Fe/H]> -1.6$. 
Unfortunately, the distribution of SX Phe in the $M_V - \log P$ 
plane does not show any significant change. 

To overcome the problem it has been suggested by Alcock et al. (2000) 
and McNamara (2000) to account for the asymmetry of the light curve, 
and indeed fundamental pulsators should present larger amplitudes and 
more asymmetric light curves when compared with first overtone ones. 
Current data for SX Phe in $\omega$ Cen seem to support this trend, 
but the observational scenario is still controversial. 
In fact, McNamara found that at least one, and maybe two, of the 
FOs in Carina present asymmetric light curves. A similar conclusion 
was also reached by Mateo, Hurley-Keller, \& Nemec (1998) 
and by Poretti (1999) on the basis of empirical data as well as 
by Bono et al. (1997) and by Templeton et al. (1998) according 
to nonlinear theoretical models. 

Finally, we decided to check whether the secular period change 
-$dP/(P\,dt)=\dot{P}/P$- 
can be adopted to overcome this thorny problem. This observable presents 
several advantages when compared with the previous ones: 1) it is the 
coupling of evolutionary and pulsational properties; 2) it is marginally 
affected by systematic errors. In particular, we estimated according to  
the pulsation relations given in \S 3.1 the variation of $\dot{P}/P$ 
along three   evolutionary tracks constructed by adopting the same 
chemical composition (Y=0.23, Z=0.001) but different stellar masses 
namely 1.2, 1.28 and 1.4 $M_\odot$.  
Figure 6 shows the change of this parameter inside the instability strip 
for the first three radial modes as a function of the logarithmic period. 
Note that the effective temperature of the blue edge was fixed according 
to the models listed in Tables 1-3, while the red edge was located 1,500 K
cooler than the previous one. The latter value is arbitrary and was 
chosen only to supply an upper limit to the width of the instability 
strip for each individual mode. 
Data plotted in this figure support the evidence that the secular 
period change can be adopted to identify the pulsation mode of SX Phe 
stars. In fact, the region in which this parameter attains similar 
values is, at fixed mode, quite narrow and located close to the blue 
edge. However, the secular period changes for the first three modes 
are quite similar for $\dot{P}/P \le 0.10$. This finding suggests 
that the mode identification is more robust for SX Phe 
located in the middle or close to the red edge of the instability strip. 
Current empirical observations on secular period changes of SX Phe are 
quite scanty and only refer to metal-rich field objects (Breger \& 
Pamyatnykh 1998).   

The photometric databases collected by microlensing experiments on 
cluster SX Phe variables can supply useful constraints on the accuracy 
of such a parameter to identify the pulsation mode. 
However, it is worth noting that to detect and measure a phase drift
in the time of maximum light due to evolutionary changes would require 
that the observations span a time interval of a few decades. In fact, 
by taking into
account a first overtone pulsator with P=5,400 s, the data plotted in the
middle panel of Fig. 6 suggest that $\dot{P}\approx3.5\times 10^{-14}s\,s^{-1}$.
Therefore according to Kepler et al. (2000) to detect a difference of a
few seconds in the observed minus calculated times of maxima is necessary
that the observations cover a time interval of $\approx10-20$ yr (see
also Breger \& Pamyatnykh 1998).

\section{High overtone models}

Together with the previous grid of models we also constructed several 
models along an evolutionary track at fixed stellar mass (\msun=1.2) 
and chemical composition (Y=0.23\footnote{The pulsational models 
distributed along the canonical track were constructed by adopting
a He content slightly higher that the evolutionary one (0.24 against 
0.23). This difference is motivated by homogeneity with previous 
calculations and also because it has a marginal effect on the pulsation 
behavior.}, Z=0.001). The adopted temperature step is roughly 300 K, 
and the modal stability was estimated up to the sixth overtone.  
We adopted a different approach to construct these models, since we 
plan to perform the nonlinear analysis on these structures.  
Moreover, to investigate the dependence of linear observables on He 
content we also constructed an evolutionary track in which, immediately  
after the central H exhaustion, the He content of the envelope was 
artificially enhanced to Y=0.30. This evolutionary track is a crude
representation of the aftermath of the collisional merging. We are 
aware that it was constructed by adopting severe over-simplifications, 
but they are only aimed at testing whether linear nonadiabatic 
observables do depend on He content. 
Figure 7 shows the two evolutionary tracks in the HR diagram together 
with the pulsation models (filled circles). 
The inner boundary for some of the hottest models was slightly moved 
toward the surface to maintain approximately constant the envelope 
mass across the strip. However in agreement with G98, we find that the 
period difference between canonical and shallower models is of the 
order of 0.0002 day and smaller than 0.0003 day for period ratios. 
Table 5 lists the input parameters, the periods and the pulsation 
constants for the two sets of models. 

Data listed in this table show that the models constructed by adopting
a higher He content present a different modal stability at hotter 
effective temperatures. In fact, among He-enhanced models the 
fundamental mode becomes unstable for temperature roughly equal 
to 7,000 K, whereas for canonical models it  becomes unstable close 
to 8,000 K. Canonical and He-enhanced models show at lower effective 
temperatures a similar behavior, and indeed for $T_e \le 7,300 K$ 
higher overtones, as expected, become rapidly stable.   

The evidence that at higher effective temperatures the He-enhanced
models present a different modal stability when compared with the 
canonical ones is worth being investigated. In order to assess 
whether this difference is caused by the increase in the He content
we performed a numerical experiment by constructing two models at 
fixed stellar mass (\msun =1.2), luminosity (\lsun =0.98), effective 
temperature ($T_e=7,900$ K),  and metal abundance (Z=0.001) but 
different helium contents, namely Y=0.24 and Y=0.30. 
Interestingly enough, we find that the pulsation properties of 
the two models are almost identical, and indeed the periods of the 
He-enhanced model are slightly longer ($<$ 1\%), while the total 
kinetic energy is larger by an amount which ranges from 15\% 
for the second overtone to less than 2\% for the sixth overtone.  
This finding suggests that at fixed input parameters an increase 
of 20\% in the He content does not cause a substantial change 
in the pulsation characteristics.  
 
As a consequence, the change in the modal stability of hot models 
is caused by evolutionary effects, and in particular by the 
increase in luminosity. Figure 8 shows the total works for the first 
six modes of the two pulsation models constructed by adopting the same 
mass value and metal content but different luminosities and He abundances  
(see labeled values). Data plotted in this figure show that the 
increase in the luminosity causes in He-enhanced models a decrease 
in the mean density, and in turn a systematic shift of both H and He 
driving regions toward the center. At the same time, the work curves 
show that the Z-bump, due to metals, plays a marginal role in the 
destabilization of higher overtones. In fact, the nonadiabatic region 
-i.e. the envelope region which drives these modes- moves toward the 
stellar surface, and therefore their properties are marginally 
affected by the innermost adiabatic regions. 

On the other hand, the periods of He-enhanced model are approximately 
20\% longer than the canonical ones (see Table 5). Moreover we also 
find that the total kinetic energy of He-enhanced model is 
$\approx$ 20\% smaller among lower modes and the difference 
becomes vanishing among higher overtones.  
This implies that SX Phe variables characterized by a large He 
content in the envelope should present larger pulsation 
destabilizations when compared with the canonical ones.  
These results, once confirmed by nonlinear, limiting cycle 
calculations, suggest that not only the period distribution but also 
the pulsation amplitudes inside the instability strip can be adopted 
to constrain the evolutionary history of SX Phe stars.  

Finally, we decided to investigate whether the secular period change 
of both He-enhanced and canonical models present a different behavior 
in the $\dot{P}/P$ vs $\log P$ plane. Figure 9 shows the secular 
period changes estimated along the two evolutionary tracks plotted 
in Figure 7. Note that the blue edges adopted to perform these 
calculations rely on the pulsation models listed in Table 5, while 
the red edges were arbitrarily fixed 1,500 K cooler than the previous 
ones. Interestingly enough, Figure 9 shows that at 
fixed period the  He-enhanced models present $\dot{P}/P$ values 
that are on average a factor of two larger than the canonical ones. 
Note that the sharp jump in the $\dot{P}/P$ values of He-enhanced 
first and second overtone models is due to the fact that the He-enhanced 
track does not present, as for the canonical one, a steady increase in 
luminosity and effective temperature. These changes are spurious and 
caused by the adjustment of the physical structure of the envelope 
soon after the change of its chemical profile. However, the sharp change
of $\dot{P}/P$ values further strengthens the strong sensitivity of 
this parameter even to small variations in the physical parameters.  
This finding supports the evidence that accurate estimates of secular 
period changes can also be adopted to constrain the physical mechanisms  
which trigger the formation of BSs.

\section{Summary and conclusions}

We constructed several sets of linear, nonadiabatic pulsation models to 
investigate the modal stability of SX Phe variables. To account for the 
pulsation behavior of SX Phe in GGCs and in dwarf galaxies we adopted 
stellar masses ranging from 1.0 to 1.4 $M_\odot$ and two different 
metal contents, namely Z=0.0001 and Z=0.001. For each selected 
mass value and chemical composition we constructed model sequences 
which cover a wide range of luminosities and effective temperatures. 

Current theoretical scenario, implemented with the more metal-rich 
calculations performed by G98, suggests that the pulsation behavior 
of SX Phe stars are partially affected by metal content. In fact, we 
find that metal-poor models present when compared with metal-rich 
ones marginal changes both in the pulsation period and in the total 
work. However, it is worth mentioning that a decrease in the 
metal content from Z=0.006 to Z=0.0001 causes a shift of the 
blue edge for the first three modes toward hotter effective 
temperatures ranging from 300 to 500 K.  
At the same time, we find that at fixed input parameters the total 
kinetic energy decreases by a factor of 5 between fundamental and first 
overtone and by a factor of 2.5 between first and second overtones.
This finding further supports the evidence suggested by G98 that 
SX Phe variables should be enthusiastic overtone pulsators, 
since a decrease in the total kinetic energy implies a decrease 
in the pulsational inertia of the envelope, and in turn an increase 
in the pulsation destabilization. 

On the basis of both current theoretical models and models 
constructed by G98 we estimated the pulsational relations 
for the first three radial modes. Interestingly enough we 
find that the inclusion of a metallicity term substantially 
improves the accuracy of the analytical periods. In fact, 
we find that over the entire period range the difference 
between computed and estimated periods is within 0.005 dex.  
Moreover and even more important, we confirm the result recently 
brought out by PCD99 that the inclusion of a metallicity term 
in the PLC relation causes a decrease in the intrinsic scatter 
of the relation. At the same time, we performed a detailed 
comparison of our PLCZ relation with a similar theoretical 
relation derived by PCD99 and with the empirical relation given 
by McNamara (1997), we find that the discrepancy over the 
entire period range is smaller than 5\%. 
This result supports the use of SX Phe variables as standard 
candles, once the pulsation mode has been properly identified.  

In order to disentangle the long-standing problem of mode 
identification we performed a detailed comparison in the $M_V$ vs
$\log P$ plane between the predicted blue edges for the three first 
radial modes and current available data for SX Phe in GGCs, in dwarf 
galaxies, and in the Galactic field. The distribution of SX Phe in the 
instability strip and current uncertainties on distance moduli 
do not allow us to properly identify the pulsation mode of 
these objects. The outcome is the same if we use the luminosity 
amplitude vs period diagram or the HR diagram. Bright, short-period, 
high-overtone pulsators as for V15 and V16 in 47~Tuc (G98) might 
provide better prospects for identification, however for these 
superior photometry from 1999 (Bruntt et al. 2000) provide a V15 
period ratio that is now inconsistent with a simple high-overtone
interpretation. For V16 the new data confirm the lower amplitude
pulsation found earlier, but the longer period mode is not detectable 
in the new data suggesting significant amplitude changes have occurred.
The high overtone mode identifications for 47~Tuc SX Phe stars have
not been shown to be secure.  

The difficulties in the mode identification can be explained 
if we assume that the instability strip of SX Phe is characterized 
by wide "OR regions" i.e. regions within which two or more modes 
are simultaneously excited. This plausible but qualitative argument  
is supported by the large number of mixed-mode pulsators found 
in 47~Tuc by G98. However, it is clear that we come to a deadlock and    
SX Phe together with Mira are the two groups of radial variables for which 
we still lack a straightforward method to single out the pulsation mode. 
Therefore it seems quite interesting that secular period changes can be 
adopted to single out the pulsation mode of cooler SX Phe variables.  
To properly attack the problem, not only new and accurate data for
SX Phe stars in GGCs and in dwarf galaxies are necessary, but also
a detailed set of full amplitude nonlinear pulsation models are needed
which can supply useful insights into their pulsation properties. 

Together with the previous calculation we also constructed two new 
sets of pulsation models along two evolutionary tracks characterized 
by the same stellar mass (1.2 $M_\odot$) and metal content but different 
He contents, namely 0.24 and 0.30. To supply a crude but straightforward 
representation of a stellar collision the latter track was constructed 
by artificially enhancing, soon after the central H exhaustion, the 
He content of the envelope. Unlike, the previous sets of models we 
estimated for these models the first seven modes i.e. fundamental 
and six overtones.  
Interestingly enough we find that the increase in the He causes, 
as expected, an increase in the stellar luminosity, which in turn  
causes a significant change in the pulsation behavior of SX Phe. 
In fact, among the He-enhanced models the fundamental mode becomes 
unstable for effective temperatures close to 7,000 K, while the 
canonical ones at 8,000 K. Moreover, the He-enhanced models present 
pulsation periods and total kinetic energies that are approximately 
20\% longer and 20\% smaller than the canonical ones.  It is noteworthy 
that the secular period changes of He-enhanced models are, on average, 
a factor of two larger than the canonical ones.  

This finding brings out the evidence that the pulsation behavior of 
SX Phe and in particular luminosity amplitudes, period distributions, 
the topology of the instability strip, and the secular period change 
can supply useful constraints on the evolutionary history of BSs, 
and in turn on the physical mechanisms which trigger their formation 
in GGCs and in dwarf galaxies.   

One of us (PS) wish to acknowledge the Rome Astronomical Observatory
for the warm hospitality during which this paper was written. 
We are grateful to an anonymous referee for his/her useful suggestions 
that served to improve the content of the paper.
This research has made use of NASA's Astrophysics Data System Abstract
Service and of SIMBAD database operated at CDS, Strasbourg, France.
This work was supported by MURST -Cofin2000- under the scientific
project: "Stellar Observables of Cosmological relevance". 

\pagebreak

\pagebreak
\figcaption {{\em Left panels:} Opacity (top) and adiabatic exponent
(bottom) for three models centrally located in the instability strip
and constructed by adopting fixed mass, luminosity, effective temperature, 
and He content but different metallicities (see labeled values).
{\em Right panels:} total work curves per logarithmic temperature for 
the three models adopted in the left panels. The individual work 
curves were normalized using the same value of the total kinetic 
energy: KE(F)=$3.92\times 10^{44}$ erg for the fundamentals (top) and
KE(FO)=$5.55\times 10^{43}$ erg for the first overtones (bottom).
Fundamental and first overtone periods (d) are also plotted in the 
right panels.}

\figcaption {Fractional difference between computed and analytical 
periods. The latter ones were estimated by adopting the pulsation 
relations which include the metallicity term. From top to bottom the 
panels refer to fundamental, first, and second overtone respectively. 
Models constructed by adopting different metal contents are plotted 
with different symbols.}   

\figcaption {Fractional difference in the bolometric magnitude at fixed 
metallicity (Z=0.006) between different fundamental PLCZ relations as 
a function of logarithmic period.  
The fractional differences -i.e. {\em (our- other)/our}- were estimated 
between our PLCZ relation and the PLCZ relations based on theoretical 
models (triangles) and empirical data (circles) for HADS stars 
(McNamara 1997) derived by PCD99. See text for more details.}   

\figcaption {Comparison in the $M_V - \log P$ plane between predicted 
blue edges for the first three radial modes and current data available 
in the literature for SX Phe in GGCs (top panel) and in the dwarf galaxies 
Carina and Sagittarius (bottom panel). See Table 4 for more details on 
empirical data. The error bar in the lower right corner is a plausible 
estimate of the uncertainties affecting current distance determinations.}  

\figcaption {Same as Figure 4, but for field SX Phe variables. Primary 
(fundamental) and secondary (first overtone) periods of double-mode 
variables are connected with a short-dashed line. } 

\figcaption {Secular period changes for the first three radial modes
as a function of the pulsation period. The period changes were estimated 
along three different evolutionary tracks constructed by adopting the 
same composition (Y=0.23, Z=0.001) but different stellar masses 
(see labeled values).}   

\figcaption{HR diagram showing the location of the pulsational
models (open circles) constructed along the evolutionary tracks 
for \msun=1.2, Y=0.23 and Z=0.001 (solid line).  
The dashed line and the triangles refer to the evolutionary track 
whose He content in the envelope was artificially enhanced to 
Y=0.30. See text for more details.} 

\figcaption {Total work curves for the first six radial modes as a 
function of the exterior mass. The two models were constructed by 
adopting the same stellar mass, effective temperature, and metal 
content (see labeled values), but different He contents and 
luminosities namely Y=0.24, \lsun=1.05 (solid line) and 
Y=0.30, \lsun=1.18 (dashed line). The individual work curves were 
normalized using the same value of the total kinetic energy (erg, 
lower right corner) and artificially enhanced by different factors 
(lower left corner). Linear periods (d) are also plotted.} 

\figcaption {Same as Figure 6, but for the evolutionary tracks 
plotted in Figure 7.}  


\begin{thebibliography}{} 
\bibitem []{} Ahumada, J., \& Lapasset, E. 1995, A\&AS, 109, 375
\bibitem []{} Alcock, C. et al. 2000, ApJ, 536, 798  
\bibitem []{} Alexander, D. R., \& Ferguson, J. W. 1994, ApJ, 437, 879  
\bibitem []{} Auriere, M., Lauzeral, C., Ortolani, S. 1990, Nature, 344, 638 
\bibitem []{} Bailyn, C. D. 1992, ApJ, 392, 519  
\bibitem []{} Bailyn, C. D. 1995, ARA\&A, 33, 133 
\bibitem []{} Bailyn, C. D., \& Pinsonneault, M. H. 1995, ApJ, 439, 705   
\bibitem []{} Benz, W., \& Hills, J. G. 1992, ApJ, 389, 546   
\bibitem []{} Bono, G., Caputo, F., Cassisi, S., Castellani, V., Marconi, M. 
\& Stellingwerf, R. F. 1997, ApJ, 477, 346 (B97)  
\bibitem []{} Bono, G., Caputo, F., Castellani, V., \& Marconi, M. 1999b, 
ApJ, 512, 711 
\bibitem []{} Bono, G., Incerpi, R., \& Marconi, M. 1996, ApJ, 467, L97  
\bibitem []{} Bono, G., \& Marconi, M. 1999, in IAU Symp. 190, New views 
of the Magellanic Clouds, ed. Chu, Y.-H., Suntzeff, N.B., Hesser, J.E., \&
Bohlender, D.A. (San Francisco: ASP), 527  
\bibitem []{} Bono, G., Marconi, M. \& Stellingwerf, R. F. 1999a, ApJS, 122, 167
\bibitem []{} Bono, G., \& Stellingwerf, R. F. 1994, ApJS, 93, 233 
\bibitem []{} Borissova, J., Catelan, M., Ferraro, F.  R., Spassova, N., 
Buonanno, R., Iannicola, G., Richtler, T. \& Sweigart, A. V. 1999, A\&A 343, 813
\bibitem []{} Breger, M., \& Pamyatnykh, A\&A, 1998, 332, 958  
\bibitem []{} Bruntt, H., et al. 2001, A\&A, submitted  
\bibitem []{} Buonanno, R., Corsi, C. E., Buzzoni, A., Cacciari, C., 
Ferraro, F. R., Fusi Pecci, F. 1994, A\&A, 290, 69 
\bibitem []{} Carretta, E., Gratton, R. G. 1997, A\&AS, 121, 95
\bibitem []{} Castelli, F., Gratton, R. G., \& Kurucz, R. L. 1997a, A\&A, 318, 841
\bibitem []{} Castelli, F., Gratton, R. G., \& Kurucz, R. L. 1997b, A\&A, 324, 432 
\bibitem []{} Deutsch, E. W., Margon, B., \& Anderson, S. F. 2000, ApJ, 530, L21
\bibitem []{} Dinescu, D. I., Girard, T. M., van Altena, W. F., Yang, T. -G., 
\& Lee, Y. -W. 1996,  AJ, 111, 1205    
\bibitem []{} Edmonds, P. D., Gilliland, R. L., Guhathakurta, P., 
Petro, L. D., Saha, A., \& Shara, M. M. 1996   
\bibitem []{} Fernie, J. D. 1992, AJ, 103, 1647  
\bibitem []{} Ferraro, F. R., Fusi Pecci, F., Bellazzini, M. 1995, A\&A, 294, 80
\bibitem []{} Ferraro, F. R., et al. 1997, A\&A, 324, 915   
\bibitem []{} Ferraro, F. R., Paltrinieri, B., Rood, R. T., Dorman, B. 1999, 
AJ, 522, 983  
\bibitem []{} Groenewegen, M. A. T. 2000, A\&A, astro-ph/0010298 
\bibitem []{} Gilliland, R. L., Bono, G., Edmonds, P. D., Caputo, F., 
Cassisi, S., Petro, L. D., Saha, A. \& Shara, M. M. 1998, ApJ, 507, 818 (G98) 
\bibitem []{} Grillmair, C. J., et al. 1998, AJ, 115, 14  
\bibitem []{} Grindlay, J. E. 1999, in Annapolis Workshop on Magnetic 
Cataclysmic Variables, ed. Hellier, C. \& Mukai, K. (San Francisco, ASP), 377  
\bibitem []{} Guarnieri, M. D., Ortolani, S., Montegriffo, P., Renzini, A., 
Barbuy, B., Bica, E., Moneti, A. 1998, A\&A, 331, 70  
\bibitem []{} Harris, W. E. 1996, AJ, 112, 1487  
\bibitem []{} Hintz, E. G., Joner, M. D., McNamara, D. H., Nelson, K. A., Moody, J. W., \& Kim , C.  1997a, PASP, 109, 15
\bibitem []{} Hintz, E., Hintz, M. L., \& Joner, M. D. 1997b, PASP, 109, 1073
\bibitem []{} H\"og, E. \& Petersen, J. O. 1997, A\&A, 323, 827   
\bibitem []{} Iglesias, C. A., \& Rogers, F. J. 1996, ApJ, 464, 943  
\bibitem []{} Kaluzny, J. 1997, A\&AS, 122, 1
\bibitem []{} Kaluzny, J., Hilditch, R. W., Clement, C., \& Rucinski, S. M. 1998, MNRAS, 296, 347 
\bibitem []{} Kaluzny, J., Krzeminski, W. 1993, MNRAS, 264, 785  
\bibitem []{} Kaluzny, J., Krzeminski, W., \& Mazur, B. 1995, AJ, 110, 2206  
\bibitem []{} Kaluzny, J., Krzeminski, W., \& Nalezyty, M. 1997, A\&AS, 125, 337
\bibitem []{} Kaluzny, J., Kubiak, M., Szymanski, M., Udalski, A., 
Krzeminski, W., Mateo, M. 1996, A\&AS, 120, 139    
\bibitem []{} Kaluzny, J., Kubiak, M., Szymanski, M., Udalski, A., 
Krzeminski, W., Mateo, M., \& Stanek, K. 1997, A\&AS, 122, 471   
\bibitem []{} Kaluzny, J., Thompson, I., Krzeminski, W., \& Pych, W. 1999, 
A\&A, 350, 496  
\bibitem []{} Kepler, S. O., et al. 2000, ApJ, 534, L185 
\bibitem []{} Kholopov, P. N., et al. 1998, GCVS, 4th Edition  
\bibitem []{} Landsman, W., Bohlin, R. C., Neff, S. G., O'Connell, R. W., 
Roberts, M. S., \& Smith, A. M., \& Stecher, T. P. 1998, AJ, 116, 789  
\bibitem []{} Lopez De Coca, P., Rolland, A., Rodriguez, E. \& Garrido, R. 1990, A\&AS, 83, 51 
\bibitem []{} Mateo, M. 1998, ARA\&A, 36, 435  
\bibitem []{} Mateo, M., Harris, H. C., Nemec, J., \& Olszewski, E. W. 1990,
AJ, 100, 469  
\bibitem []{} Mateo, M., Hurley-Keller, D., \& Nemec, J. N. 1998, AJ, 115, 1856 
\bibitem []{} Mateo, M., Udalski, A., Szymanski, M., Kaluzny, J., 
Kubiak, M., Krzeminski, W. 1995, AJ, 109, 588   
\bibitem []{} Mazur, B., Kaluzny, J., \& Krzeminski, W. 1999a, MNRAS, 306, 727 
\bibitem []{} Mazur, B., Krzeminski, W., \& Kaluzny, J. 1999b, AcA, 49, 551  
\bibitem []{} McNamara, D. H. 1995, AJ, 109, 1751 
\bibitem []{} McNamara, D. H. Powell, J. M., \& Joner, M. D. 1996, PASP, 108, 1098 
\bibitem []{} McNamara, D. H. 1997, PASP, 109, 1221 
\bibitem []{} McNamara, D. H. 2000, PASP, 112, 1096   
\bibitem []{} Mochejska, B. J., Kaluzny, J., \& Thompson, I. 2000, astro-ph/0002341 
\bibitem []{} Monkiewicz, J., et al. 1999, PASP, 111, 139  
\bibitem []{} Nemec, J. M., Nemec-Linnell, A. F., \&  Lutz, T. E. 1994, 108, 222
\bibitem []{} Nemec, J. M. \& Mateo, M. 1990, in Confrontation between 
stellar pulsation and evolution, ed. Cacciari, C. \& Clementini, G. 
(San Francisco: ASP), 64
\bibitem []{} Nemec, J. M. \& Park, N. 1996, in The origins, evolution, and 
destinies of binary stars in clusters, ed. Milone, E.F. \& Mermilliod, J.-C.
(San Francisco: ASP), 359  
\bibitem []{} Niss, B. 1981, A\&A, 98, 415  
\bibitem []{} Ouellette, J. A., Pritchet, C. J. 1998, AJ, 115, 2539   
\bibitem []{} Petersen, J. O., \& Christensen-Dalsgaard, J. 1996, A\&A, 312, 463
\bibitem []{} Petersen, J. O., \& Christensen-Dalsgaard, J. 1999, A\&A, 352, 547 (PCD99)   
\bibitem []{} Petersen, J. O., \& H\"og, E. 1998a, A\&A, 331, 989   
\bibitem []{} Petersen, J. O., \& H\"og, E. 1998b, MemSait, 69, 59 
\bibitem []{} Piotto, G., Zoccali, M., King, I. R., Djorgovski, S. G., 
Sosin, C., Dorman, B., Rich, R. M., \& Meylan, G. 1999, AJ, 117, 264  
\bibitem []{} Poretti, E. 1999, A\&A, 343, 385  
\bibitem []{} Rodr\'iguez, E., \& L\`opez-Gonz\'alez, M. J. 2000, A\&A, 359, 957 (RLG00)  
\bibitem []{} Rodriguez, E., Lopez De Coca, P., Rolland, A. \& Garrido, R. 1990,
RMxA\&A 20, 37 
\bibitem []{} Rodriguez, E., Rolland, A., Lopez de Coca, P., Garcia-Lobo, E., 
Sedano, J. L. 1992, A\&AS, 93, 189  
\bibitem []{} Rubenstein, E. P., \& Bailyn, C. D. 1996, AJ, 111, 260  
\bibitem []{} Sandquist, E. L., Bolte, M., Hernquist, L. 1997, 
ApJ, 477, 335 (SBH97)  
\bibitem []{} Saviane, I., Piotto, G., Fagotto, F., Zaggia, S., 
Capaccioli, M., \& Aparicio, A. 1998, A\&A, 333, 479 
\bibitem []{} Shara, M. M., Saffer, R. A., \& Livio, M. 1997, ApJ, 489, L59 
\bibitem []{} Sigurdsson, S., Davies, M. B., \& Bolte, M. 1994, ApJ,  431, L115 
\bibitem []{} Sills, A., Bailyn, C. D. 1999, ApJ, 513, 428  
\bibitem []{} Templeton, M. R., Guzik, J. A., \& McNamara, B. J. 1998, BAAS, 193, 6405  
\bibitem []{} Thompson, I. B., Kaluzny, J., Pych, W., Krzeminski, W. 1999, AJ, 118  
\bibitem []{} Walker, A. R., \& Nemec, J. M. 1996, AJ, 112, 2026  
\bibitem []{} Wehlau, A., Rucinski, S. M., Shi, J., Fahlman, G. G., \& Thompson, I. 1996, IBVS, 4394, 1  
\bibitem []{} Wilson, W. J. F., Milone, E. F., Fry, D. J. I., \& van Leeuwen, J.1998, PASP, 110, 433  
\bibitem []{} Yan, L., \& Reid, I. N. 1996, MNRAS, 279, 751   

\end{thebibliography}
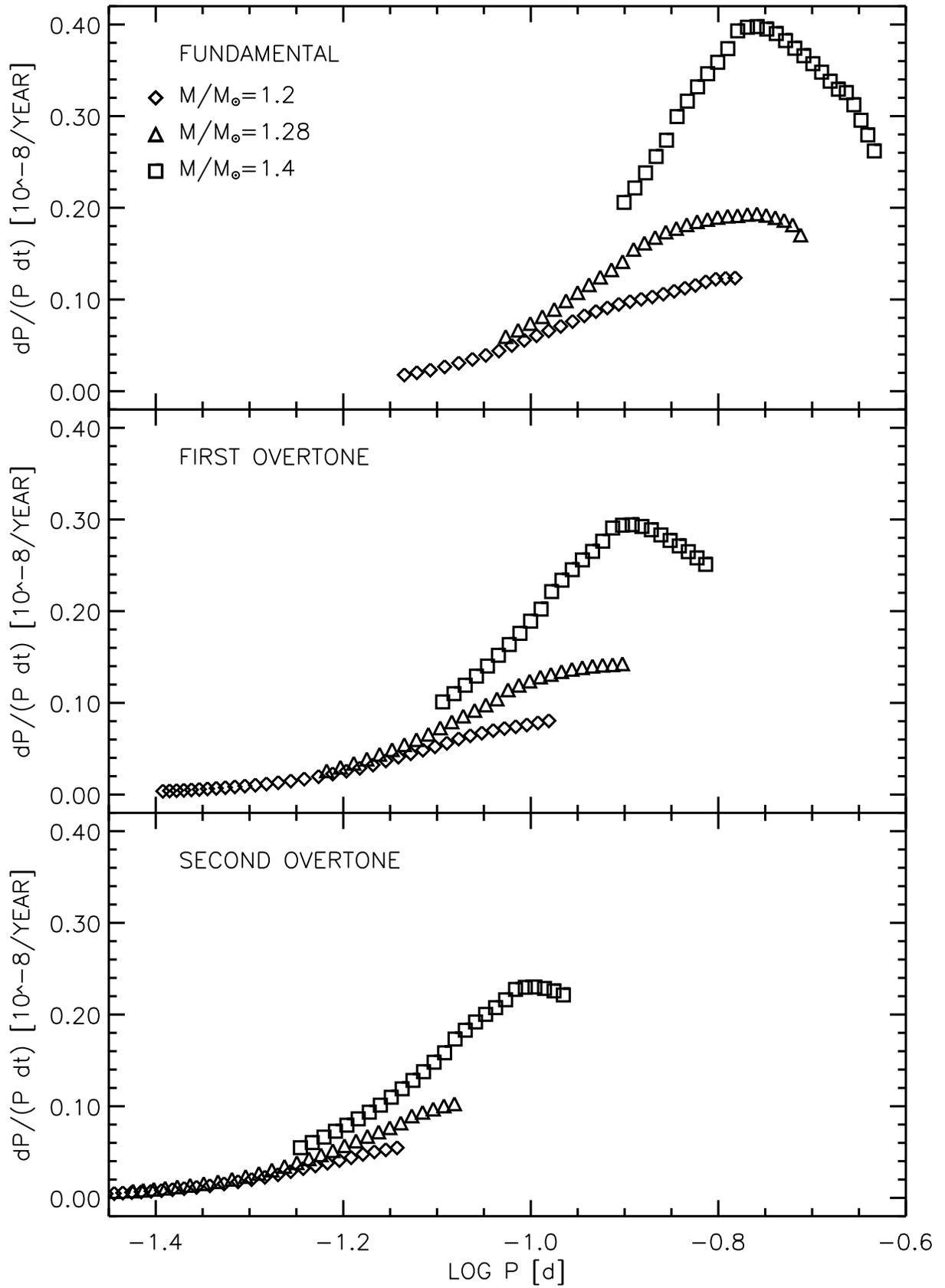
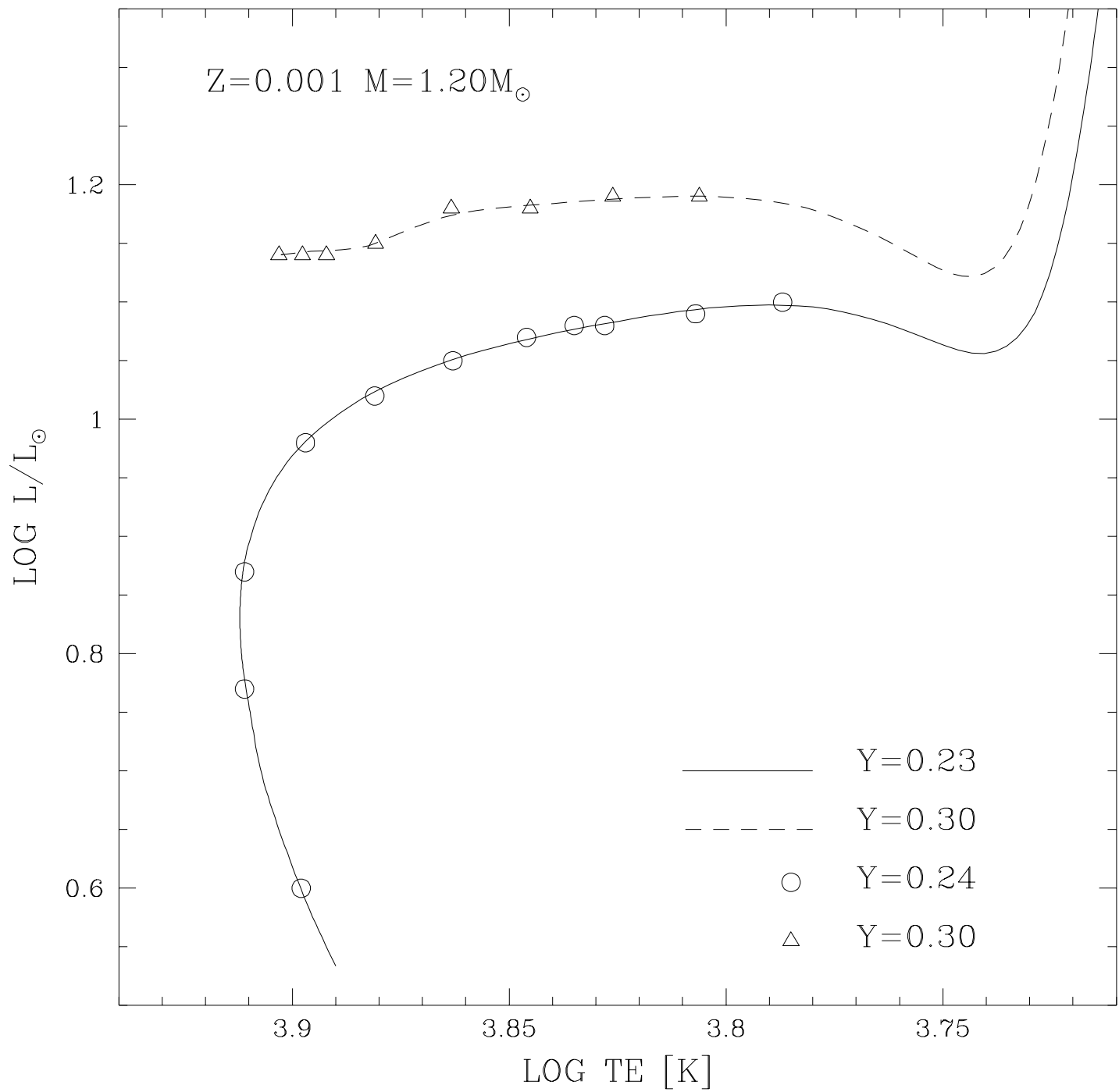
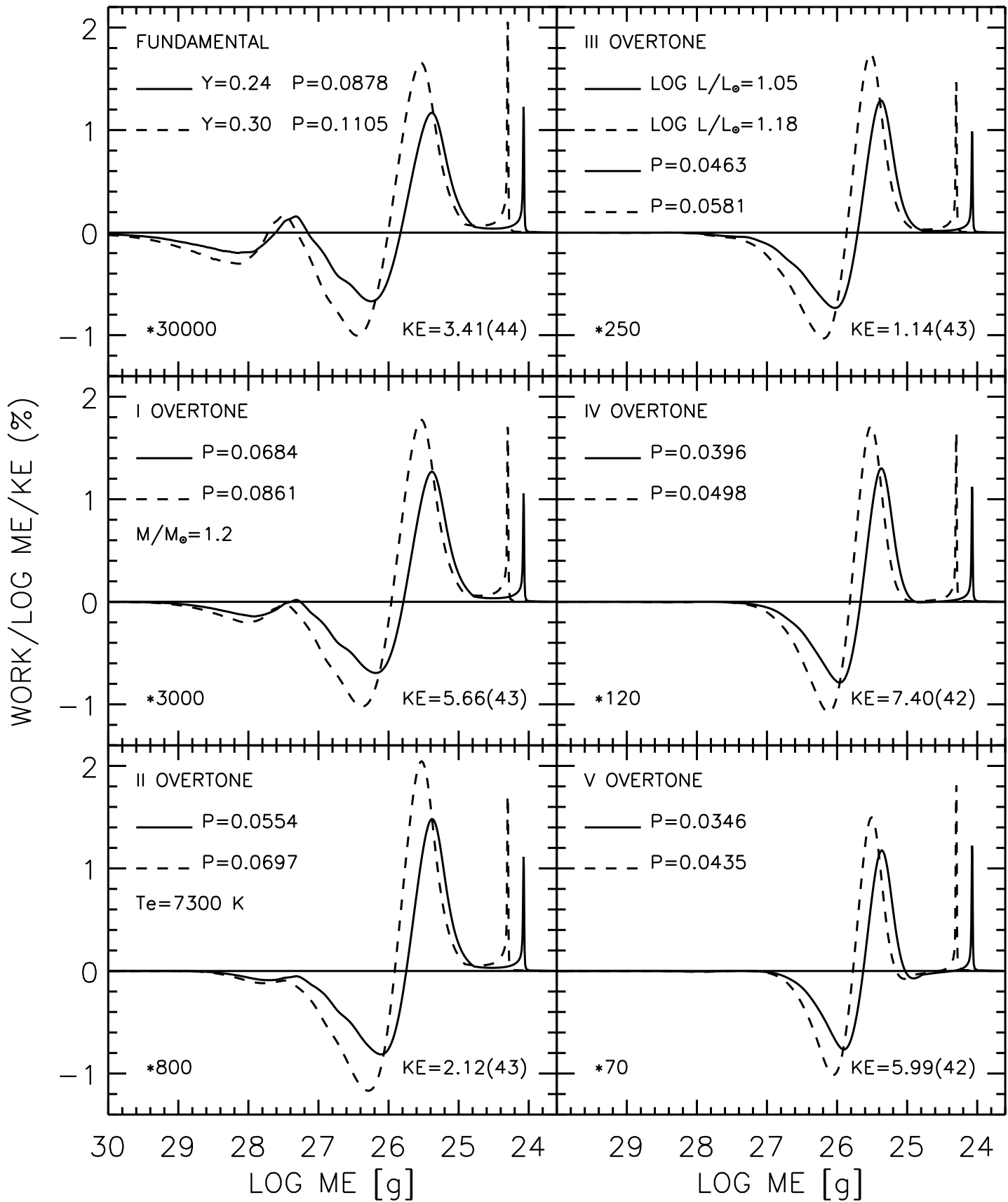
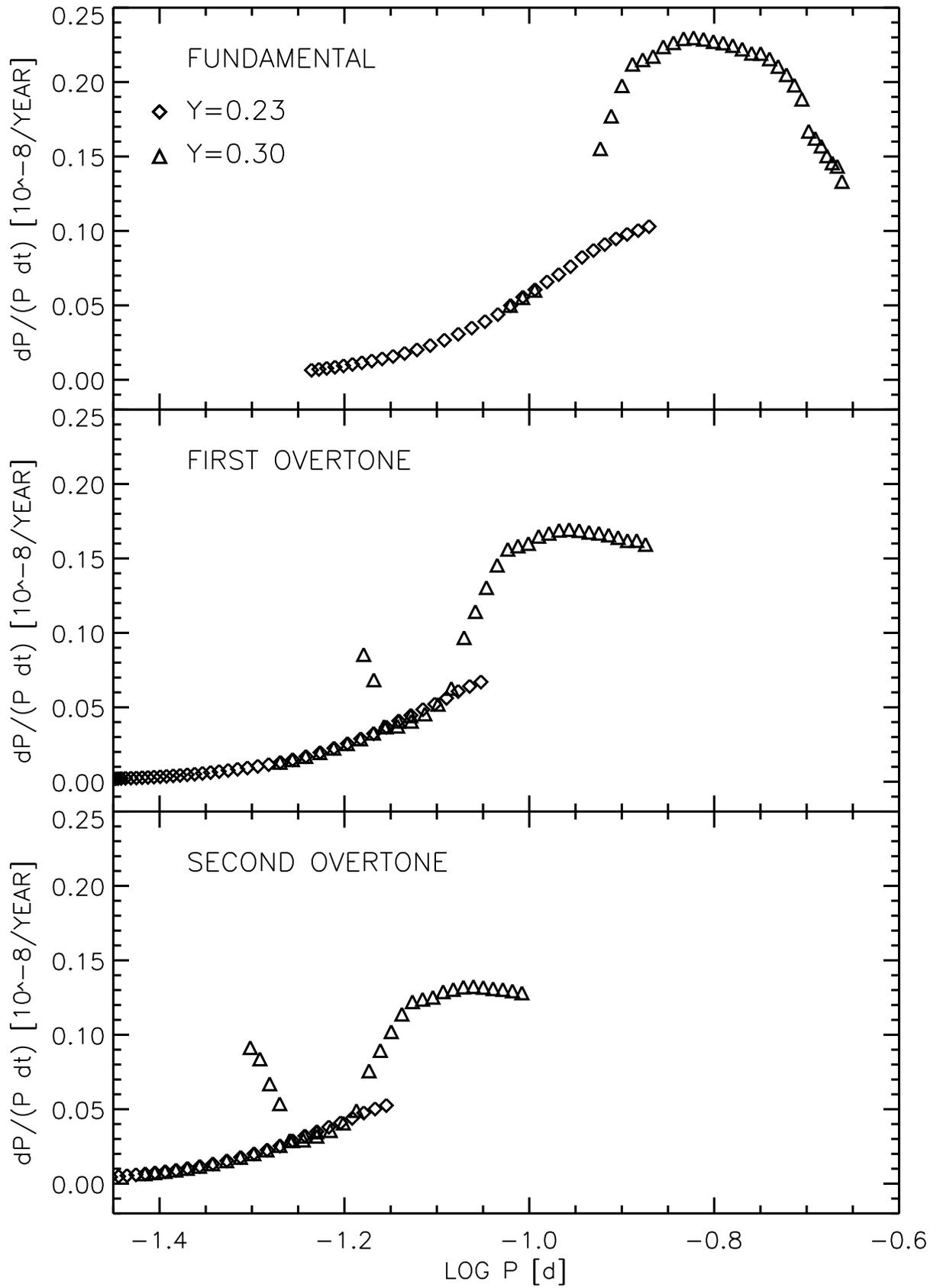
\end{document}